\documentclass[aps,prl,twocolumn,showpacs,superscriptaddress]{revtex4-1}
\usepackage{graphicx}

\begin{document}

\title{Reconciling monolayer and bilayer $J_{\rm eff} = 1/2$ square lattices in hybrid oxide superlattice}
\author{Dongliang Gong}
\affiliation{Department of Physics and Astronomy, University of Tennessee, Knoxville, Tennessee 37996, USA}
\author{Junyi Yang}
\affiliation{Department of Physics and Astronomy, University of Tennessee, Knoxville, Tennessee 37996, USA}
\author{Lin Hao}
\affiliation{Anhui Key Laboratory of Condensed Matter Physics at Extreme Conditions, High Magnetic Field Laboratory, HFIPS, Anhui, Chinese Academy of Sciences, Hefei, 230031, China}
\author{Lukas Horak}
\affiliation{Department of Condensed Matter Physics, Charles University, Ke Karlovu 3, Prague 12116, Czech Republic}
\author{Evguenia Karapetrova}
\affiliation{Advanced Photon Source, Argonne National Laboratory, Argonne, Illinois, 60439, USA}
\author{Joerg Strempfer}
\affiliation{Advanced Photon Source, Argonne National Laboratory, Argonne, Illinois, 60439, USA}
\author{Yongseong Choi}
\affiliation{Advanced Photon Source, Argonne National Laboratory, Argonne, Illinois, 60439, USA}
\author{Jong-Woo Kim}
\affiliation{Advanced Photon Source, Argonne National Laboratory, Argonne, Illinois, 60439, USA}
\author{Philip J. Ryan}
\affiliation{Advanced Photon Source, Argonne National Laboratory, Argonne, Illinois, 60439, USA}
\author{Jian Liu}
\affiliation{Department of Physics and Astronomy, University of Tennessee, Knoxville, Tennessee 37996, USA}
\email{jianliu@utk.edu}

\begin{abstract}

The number of atomic layers confined in a two-dimensional structure is crucial for the electronic and magnetic properties. Single-layer and bilayer $J_{\rm eff} = 1/2$ square lattices are well-known examples where the presence of the extra layer turns the XY-anisotropy to the $c$-axis anisotropy. We report on experimental realization of a hybrid SrIrO$_3$/SrTiO$_3$ superlattice that integrates monolayer and bilayer square lattices in one layered structure. By synchrotron x-ray diffraction, resonant x-ray magnetic scattering, magnetization, and resistivity measurements, we found that the hybrid superlattice exhibits properties that are distinct from both the single-layer and bilayer systems and cannot be explained by a simple addition of them. In particular, the entire hybrid superlattice orders simultaneously through a single antiferromagnetic transition at temperatures similar to the bilayer system but with all the $J_{\rm eff} = 1/2$ moments mainly pointing in the $ab$-plane similar to the single-layer system. The results show that bringing monolayer and bilayer with orthogonal properties in proximity to each other in a hybrid superlattice structure is a powerful way to stabilize a unique state not obtainable in a uniform structure.

\end{abstract}

\maketitle

The search for new emergent states has been the frontier of quantum materials research and often achieved by tuning competing or cooperative interactions among the spin, charge, orbital, and lattice degrees of freedom \cite{giustino20212021,tokura2017emergent,imada1998metal}. Dimensionality is a unique and critical control parameter for this purpose, since quantum confinement may enhance electronic correlation \cite{wehling2011strength,huang2017layer}, lead to stronger fluctuations \cite{mermin1966absence,coleman1973there}, and trigger exotic phenomena beyond mean-field theory \cite{imada1998metal,davies1998physics}. While the effects of dimensionality can be revealed by driving a system, for example, from the two-dimensional (2D) limit to the three-dimensional (3D) limit or the other way around, drastic changes do not necessarily take the full crossover to arise. Distinct behaviors and phases could readily occur when one more atomic plane is added to a 2D quantum confinement structure, e.g., going from a monolayer system to a bilayer system \cite{lee2006doping,tokura2000colossal,grigera2001magnetic,wu2020electronic,spaldin2005renaissance}.

One of the best examples of such dimensionality-control is the Ruddlesden-Popper (RP) series $A$$_{n+1}$$B$$_n$$X$$_{3n+1}$, where $A$ and $B$ are cations, $X$ is an anion, and $n$ controls the number of $BX{_6}$ octahedral layers. For instance, in layered ruthenates Sr$_{n+1}$Ru$_n$O$_{3n+1}$, the single-layer Sr$_2$RuO$_4$ ($n$ = 1) is an unconventional superconductor with potential $p$-wave pairing  \cite{maeno2001intriguing,wu2020electronic}, whereas the bilayer Sr$_3$Ru$_2$O$_7$ ($n$ = 2) is a paramagnetic metal with competing ferromagnetic and antiferromagnetic spin fluctuations \cite{capogna2003observation}, and it achieved a quantum critical point (QCP) by tuning the magnetic field \cite{grigera2001magnetic,lester2015field}. In layered manganese oxide La$_{n(1-x)}$Sr$_{n(x+1)}$Mn$_n$O$_{3n+1}$, the bilayer ($n$ = 2) manganites exhibit colossal magnetoresistance (CMR) \cite{rao1996giant}, but the single-layer La$_{1+x}$Sr$_{1-x}$MnO$_4$ does not despite a rich doping phase diagram \cite{moritomo1995magnetic,moritomo1996giant,sternlieb1996charge,larochelle2005structural}. Abrupt electronic and magnetic changes between monolayer and bilayer were also found in other families, such as lanthanum cuprates and nickelates \cite{lee2006doping,koster2015epitaxial}. While the monolayer-bilayer contrast has been highlighted in many quantum material families, what will happen if they are brought in proximity to each other to form a monolayer-bilayer double quantum well remains an interestingly open question. However, it is generally difficult to realize such a hybrid structure in real materials.

\begin{figure}
\includegraphics[scale=0.6]{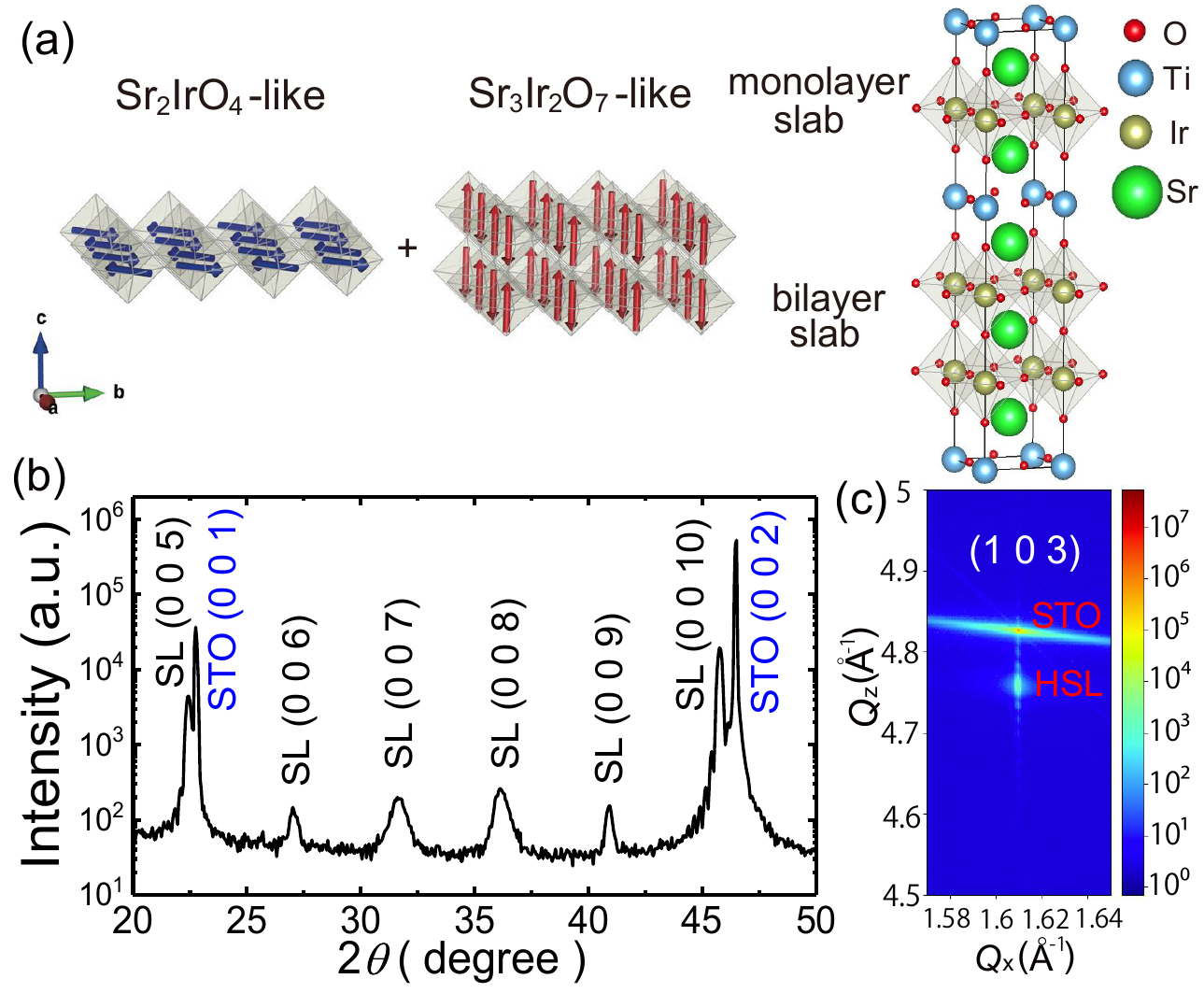}
\caption{ (a) The schematic diagram of monolayer and bilayer with orthogonal spin symmetries combined in a HSL structure. The right side shows a structural schematic of HSL, including alternative SrIrO$_3$ (olive) and SrTiO$_3$ (dark blue) layers. (b) X-ray diffraction pattern of the HSL film along (0 0 L) direction at room temperature. (c) RSM around the SrTiO$_3$ substrate (103) peak.}
\label{fig1}
\end{figure}

Here we show that oxide heteroepitaxy could provide an attractive solution in virtual of the layer-controlled synthesis of quantum well structures. We demonstrated this approach through an artificial superlattice (SL) where we combined monolayer and bilayer $J_{\rm eff} = 1/2$ square lattices by stacking one- and two-unit-cell thick SrIrO$_3$ slabs with SrTiO$_3$ spacing layers in-between [Fig.1 (a)]. Such a layered superstructure is an ideal prototype because a dimensionality-controlled spin-flop transition between monolayer and bilayer iridates have been well established \cite{bertinshaw2019square,ye2013magnetic,boseggia2012magnetic,murayama2021bond,hao2017two,yang2020strain,meyers2019magnetism,seyler2020spin,fujiyama2012weak,boseggia2012antiferromagnetic}. Specifically, while both of them have the $J_{\rm eff} = 1/2$ antiferromagnetic (AFM) insulating state cooperatively stabilized by the strong spin-orbit coupling and the electronic correlation \cite{bertinshaw2019square,kim2014fermi,murayama2021bond,cao2016hallmarks,zhao2016evidence,yan2015electron,hu2019strong}, the XY-anisotropy of the monolayer is turned into the strong $c$-axis Ising anisotropy in the bilayer due to the interlayer coupling \cite{kim2008novel,kim2012dimensionality,kim2012large,mazzone2021laser,jackeli2009mott} [Fig.1 (a)]. Despite the debate on the origin of the interlayer coupling and its role in the spin-flop transition \cite{kim2012dimensionality,kim2017dimensionality,mohapatra2019octahedral,suwa2021exciton,mazzone2022antiferromagnetic,meyers2018decoupling}, bringing the monolayer and the bilayer together in one structure would create an intriguing scenario to explore the proximity effect of two states with orthogonal anisotropy. The enforced competition or cooperation between monolayer and bilayer makes such a hybrid structure a good candidate to achieve new emergent states \cite{witczak2014correlated,wang2011twisted,meng2014odd,jackeli2009mott,carter2012semimetal,chen2014topological}.

In this Letter, we report our successful synthesis of a hybrid SL (HSL) of [(SrIrO$_3$)$_1$/(SrTiO$_3$)$_1$/(SrIrO$_3$)$_2$/(SrTiO$_3$)$_1$] that is designed to combine the supercells in the single-layer SL of [(SrIrO$_3$)$_1$/(SrTiO$_3$)$_1$] and the bilayer SL of [(SrIrO$_3$$)$$_2$/(SrTiO$_3$)$_1$] as shown in the Fig. 1(a). Such a hybrid structure can be considered as $n = 1.5$  member of the RP series that is analogue to Sr$_5$Ir$_3$O$_{11}$ but without the in-plane half-unit-cell slide between the monolayer and bilayer slabs \cite{harlow1995effects}. We found that the HSL shows an insulating ground state with a single AFM transition at 120 K which is lower than the single-layer SL (150 K) but similar to the bilayer SL \cite{matsuno2015engineering,hao2017two}. On the other hand, the HSL show a significantly larger net canted moment than the bilayer SL. The resonant x-ray magnetic scattering (RXMS) measurement reveals that the AFM moments are not only predominantly in-plane in the monolayer slab but also stabilized in the bilayer slab. These results show that the HSL forces the monolayer and bilayer to merge electronically and magnetically to form a new rounded system that behaves distinctively.

The HSL samples were fabricated on (001)-oriented SrTiO$_3$ (STO) substrates by pulsed laser deposition with $in$ $situ$ reflection high energy electron diffraction (RHEED). The details of the growth conditions can be found in Ref.\cite{hao2017two,hao2019anomalous,yang2020strain,hao2018giant}. The HSL consists of 20 hybrid supercells by monitoring the intensity of RHEED patterns. Single-crystal x-ray diffraction (XRD) measurements were carried out on a Panalytical X'pert MRD diffractometer with wavelength of 1.54 \r{A} at room temperature. Synchrotron XRD measurements and RXMS measurements were performed at the 33-BM-C and 4-ID-D beamlines, respectively, at the Advanced Photon Source of the Argonne National Laboratory. The in-plane resistivity was measured by the standard four-probe method with the Physical Property Measurement System from Quantum Design. The magnetization measurements were performed using the Magnetic Property Measurement System with the Vibrating Sample Magnetometer mode from Quantum Design.

\begin{figure}
\includegraphics[scale=0.32]{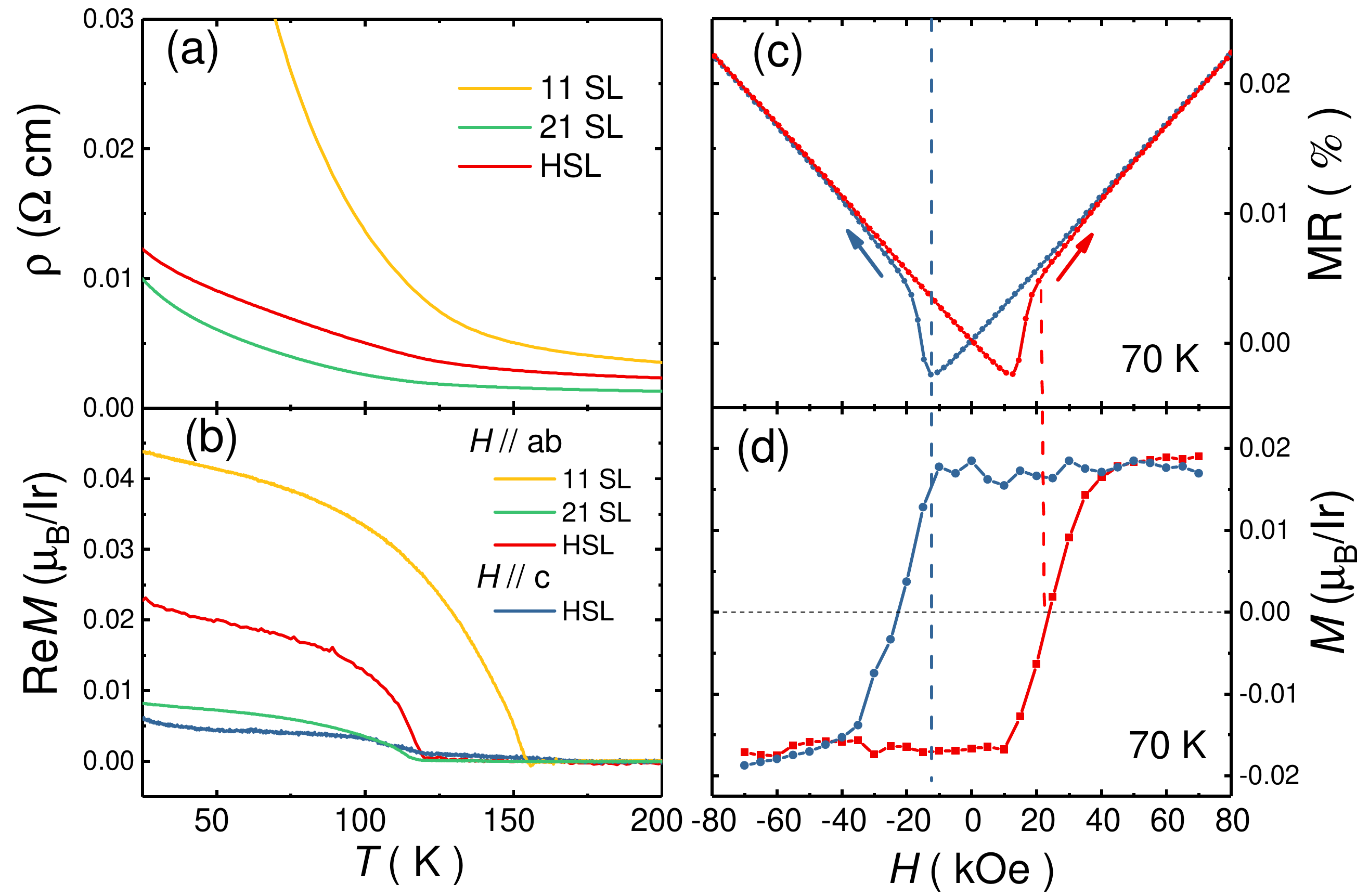}
\caption{ (a)Temperature dependence of the resistivity. The data of 11 and 21 SL are from ref.\cite{hao2017two} (b)Temperature dependence of the remnant magnetization. The data of 11 and 21 SL are from ref.\cite{hao2017two} The remnant magnetization was measured with warming up under Zero field after cooling down to 5 K with 5 kOe. (c) and (d) show the magnetic field dependence of magnetoresistance and magnetization at 70 K, respectively. The arrow marked the field sweep direction.}
\label{fig2}
\end{figure}

Figure 1(b) shows the $\theta$-2$\theta$ scan along the (0 0 $L$) direction. In addition to the perovskite peaks, we found the five SL peaks in one diffraction period of the perovskite structure, consistent with the fact that the HSL is five times of a pseudocubic perovskite cell along the $c$-axis [see Fig.1(a) ]. The indexes of the HSL are thus defined by the hybrid supercell $a \times a \times 5c$ ($a$ = 3.905 \r{A}, 5$c$ = 19.77 \r{A}). The well-defined Kiessig fringes suggest sharp substrate-film interfaces and flat surfaces, indicative of high-quality epitaxial growth. Figure 1(c) shows the reciprocal space mapping (RSM) that demonstrates the film hosts the same in-plane pseudocubic lattice parameter as the substrate, confirming a fully strained state.

Figure 2(a) show the temperature-dependent resistivity, which indicates an insulating ground state of the HSL. This insulating behavior in comparison is slightly stronger than the bilayer SL but clearly weaker than the single-layer SL, suggesting that the conduction of the HSL cannot be simply viewed in a parallel-resistor-like picture where the monolayer and bilayer slabs host conduction completely independently. In other words, the finite tunneling across the STO spacer necessarily contributes to the total bandwidths and the final gap sizes of all the SLs \cite{kim2016manipulation,hao2017two}. Note that the HSL is different from both the monolayer and bilayer SLs in that the charge is now hopping between two dissimilar slabs. If the electronic structures of the two slabs are non-degenerate and they have different individual gap sizes, the inter-slab hybridization of the HSL will end up being suppressed from that of a uniform SL, which could account for the observed intermediate strength of the insulating behavior \cite{autieri2014structural}.

We also measured the temperature-dependent remnant magnetization which provides a convenient probe of the AFM transition in many iridates in virtue of the canted moment induced by octahedral rotation. The data in Figure 2(b) reveals a magnetic transition of the HSL at around 120 K, which is close to the transition temperature of the bilayer SL and much lower than the single-layer SL \cite{hao2017two,matsuno2015engineering}. The size of the in-plane net moment per Ir site is however much bigger than that of the bilayer SL and about half of the single-layer SL \cite{hao2017two}. These results have a number of important implications. Firstly, it suggests that the magnetic ground state of the HSL is a canted AFM order. Secondly, the observed predominant in-plane canted moments and the small but finite out-of-plane canted moments require the AFM moments of at least one of the two slabs to have an in-plane component. If only one of them has planar moments, it is most likely to be the monolayer slab given its own reported XY-anisotropy \cite{hao2019anomalous,yang2020strain}. In the other extreme, if both slabs have and only have planar moments, the canting angle and/or the averaged size of the local ordered moments would have to be considerably reduced from that of the single-layer SL, unless the two slabs have opposite canting directions. We ruled out the possibility of having opposite canting directions by magnetic hysteresis loop measurements of magnetoresistance and magnetization, both of which shown in Figure 2(c) and (d) exhibit a single coercivity field consistently. Last but not least, regardless of the anisotropy, a single AFM transition of the HSL would mean that both slabs are ordered simultaneously. In other words, since the inter-slab coupling is critical to the long-range order stability of such a quasi-2D system \cite{hao2017two,hao2018giant}, the monolayer slab in the HSL cannot order without the adjacent bilayer slabs being ordered as well, accounting for the observed transition temperature. In short, what is already clear from magnetometry is that the HSL is not just a simple superposition of the monolayer and bilayer.

\begin{figure}
\includegraphics[scale=0.55]{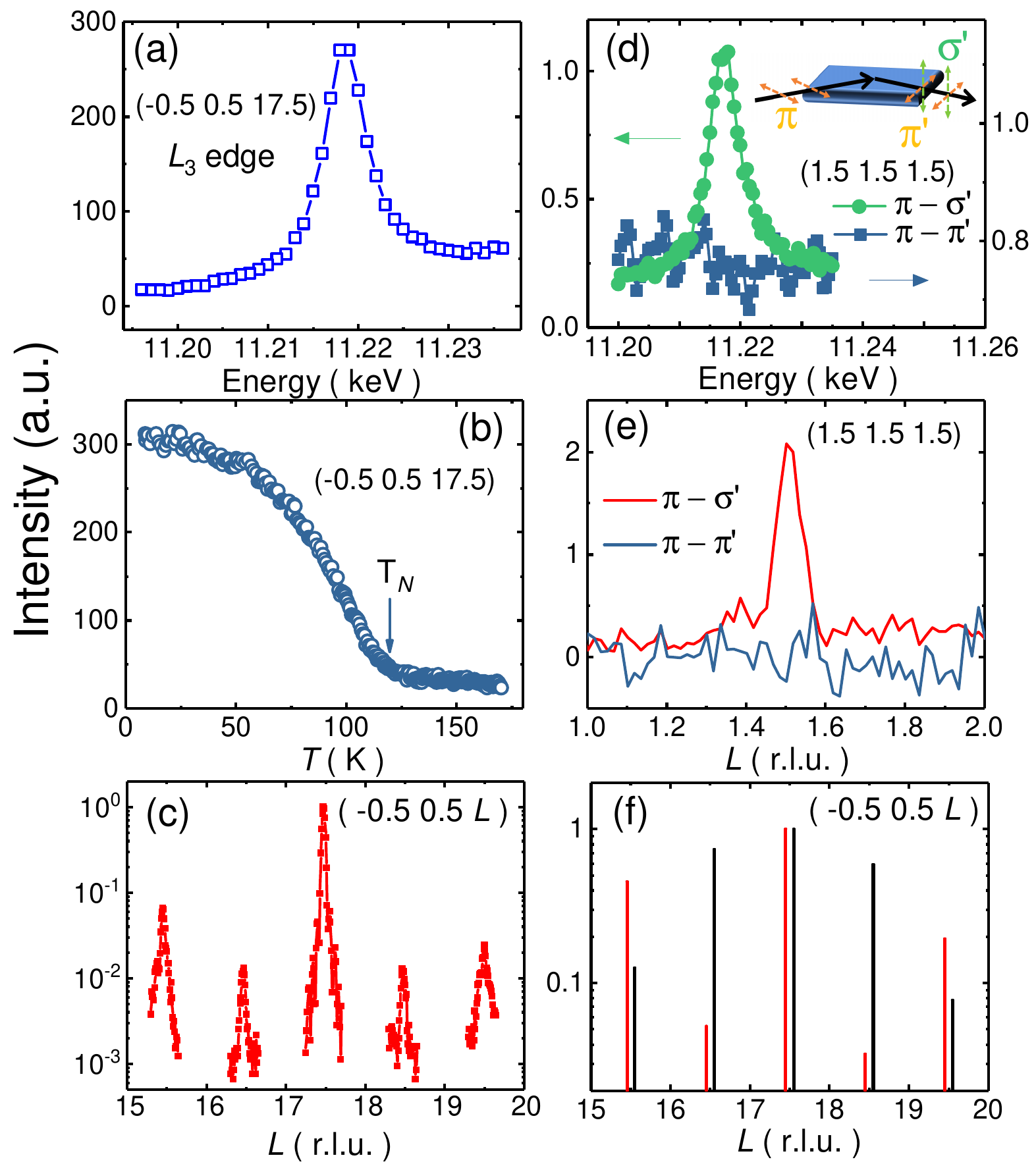}
\caption{ (a) Energy scan at the (-0.5 0.5 17.5) magnetic reflection near the Ir $L_3$ edge at 8 K. (b) Temperature dependent intensity of the (-0.5 0.5 17.5) peak. (c) $L$ scans of magnetic diffraction along the [-0.5, 0.5, $L$] direction. (d) Energy scans at (1.5 1.5 1.5) in $\pi - \pi ^{\prime}$ and $\pi - \sigma ^{\prime}$ channel. (e) $L$ scan of magnetic diffraction along $L$ direction in $\pi - \pi ^{\prime}$ and $\pi - \sigma ^{\prime}$ channel. The intensity is calculated from the intensity difference at 11.216 and 11.20 keV. (f) Calculated $L$ dependence of intensity along [-0.5 0.5 $L$]. The black line is calculated with moments along $c$-axis in bilayer slab, and red line is calculated with moments in $ab$-plane.They are shifted $\Delta L = \pm$ 0.1 for clarity.  }
\label{fig3}
\end{figure}

To unambiguously distinguish the different scenarios discussed above, it is vital to resolve the AFM structure of the HSL, especially the moment orientation. However, directly probing the AFM structure and transition is challenging as the limited volume of ultrathin film samples precludes many methods applicable to the bulk, such as neutron scattering \cite{ye2013magnetic,ye2015structure}. RXMS offers an ideal solution thanks to its resonant enhancement and high photon energy in the hard x-ray regime. Figure 3(a) shows the spectrum of the (-0.5 0.5 17.5) peak observed under the vertical scattering geometry in the $\sigma - \pi ^{\prime}$ channel, which shows a typical magnetic line shape at the Ir $L_3$-edge \cite{hao2017two,meyers2019magnetism,kim2009phase,boseggia2013robustness,ohgushi2013resonant}. Magnetic peaks at half order of $H$ and $K$ are characteristic of a nearest-neighbor AFM order within the $ab$-plane, consistent with the expected magnetic ground state of a pseudospin-half square lattice. Figure 3(b) shows the temperature dependence of the intensity, which reveals the transition temperature $\it{T}$$_N$ = 120 K. This is also the only observed transition, consistent with the magnetization result. Interestingly, including (-0.5 0.5 17.5), magnetic peaks are only found at the half-order positions of $L$ as shown in Fig.3(c), suggesting that the magnetic cell consists of two hybrid supercells that are magnetically out-of-phase from each other along the $c$-axis. To probe the AFM moment direction, we measured the (1.5 1.5 1.5) magnetic peak under the horizontal geometry with polarization analysis. Figure 3(d) and (e) show that the peak only has intensity in the $\pi - \pi ^{\prime}$ channel but no intensity in the $\pi - \sigma ^{\prime}$ channel, suggesting that the AFM moments of the entire HSL are predominantly in-plane and the $c$-axis component is weak (if not zero) \cite{calder2015evolution,kim2022sr}. 

The most important implication of this result is that the AFM moments of the two slabs are not orthogonal unlike those in their individual systems \cite{hao2017two,yang2020strain,meyers2019magnetism,matsuno2015engineering}. To confirm this, we simulated the $L$-modulation of the (-0.5 0.5 $L$) magnetic peaks \cite{supp} because the relative change in the magnetic structural factor at different $L$ positions is primarily driven by the included angle of the AFM moments between the monolayer and bilayer slabs. We assumed the same moment sizes for both slabs and compared two simple scenarios: (i) AFM moments in $ab$-plane for monolayer and along $c$-axis for bilayer; (ii) AFM moments in $ab$-plane for both monolayer and bilayer with a parallel alignment. The former has a 90$^{\circ}$ included angle, while the latter's is zero. Comparing these two scenarios not only captures the impact of the included angle but also takes into account the in-plane moments in at least one slab according to the magnetization data. As can be seen in Fig.3(f), scenario (ii) clearly reassembles the observed modulation much better than (i). In particular, the intensity at $L= 17.5$ is strongest in (ii), whereas $L= 16.5$ and 18.5 are the two largest ones in (i), in sharp contrast to the experiment. Their distinct behaviors is due to the fact that the form factors of the two slabs are destructively superimposed at $L= 16.5$ and 18.5 and constructively added up at $L= 17.5$. Therefore, relative to the peak at $L= 15.5$ where the superposition is neither constructive nor destructive, having a strong peak at $L= 17.5$ and a weak peak at $L= 16.5$ is a signature that the two slabs have similar form factors and hence a small included angle. While including more variables and subtle modifications to the magnetic structure may further improve the agreement of the simulation with the experiment \cite{supp}, this analysis demonstrates that the HSL cannot be regarded as a simple addition of the two subsystems.

\begin{figure}
\includegraphics[scale=0.35]{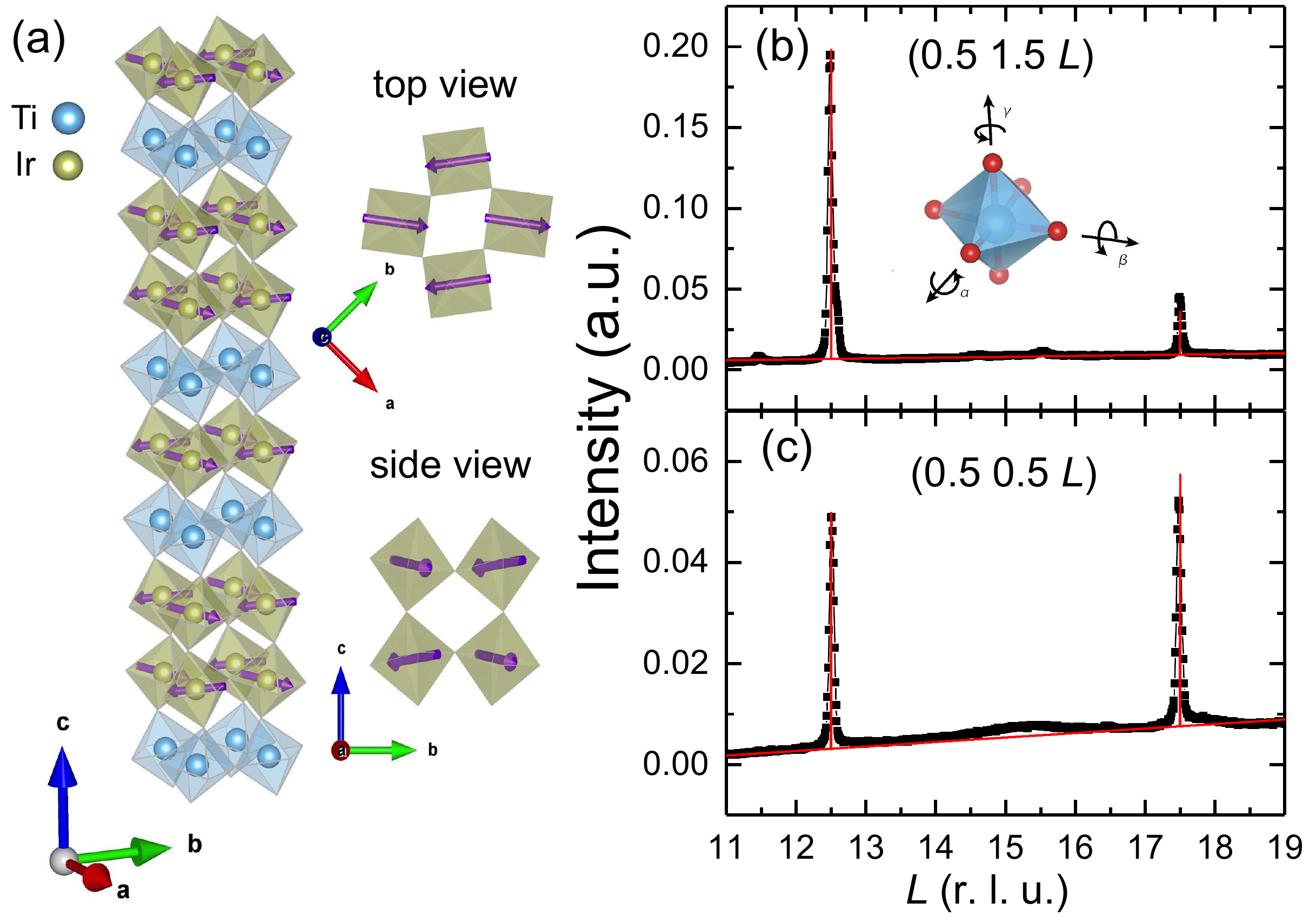}
\caption{ (a) Magnetic structure of 1121 SL. The purple arrows indicate the moment directions. Half-order peaks at (0.5 1.5 $L$) (b) and (0.5 0.5 $L$) (c) arising from the octahedral rotation and tilt. The inset in (b) shows the rotation axes in perovskite. The red line is calculated with 9$^{\circ}$ rotation angle and 4$^{\circ}$ tilt angle \cite{supp}.}
\label{fig4}
\end{figure}

Figure 4(a) shows the complete magnetic structure of the HSL considering the AFM order in scenario (ii) and the finite net moment from the spin canting. In particular, since the in-plane AFM moments are out-of-phase between the two hybrid supercells inside the magnetic cell, the only way for their canted moments to be in-phase is to have the octahedra rotated out-of-phase as well from one hybrid supercell to another around the $c$-axis. This could be achieved by a perovskite $c^{-}$ rotation (Glazer's notation) \cite{glazer1975simple}, given the total odd number of perovskite layers in the hybrid supercell. We verified this structural distortion by measuring the corresponding half-order structural peaks at (0.5 1.5 12.5) and (0.5 1.5 17.5) as shown in Fig. 4(b). No peak associated with $c^{+}$ rotation was detected. We estimated the rotation angle ($\gamma$) to be about 9$^{\circ}$ which is similar to the single-layer and bilayer SLs \cite{supp,matsuno2015engineering}. This suggests the canting angle is also similar to the individual SLs, which indicates that the reduced canted moment of the HSL compared with the single-layer SL is mainly caused by a decrease of the local ordered moment size. A reduced order parameter signifies weaker charge localizations or stronger charge fluctuations in the HSL due to a relatively larger effective bandwidth \cite{hao2019anomalous,yang2020strain}, which is actually consistent with the resistivity data as well as a smaller $\it{T}$$_N$.

In addition to the rotation peaks, we also measured the half-order structural peaks associated with the octahedral tilting around the $a$- and $b$-axes, since buckling the vertical Ir-O-Ir bond has been shown to decrease the bilayer $c$-axis anisotropy \cite{meyers2019magnetism,mohapatra2019octahedral}. As shown in Fig.4(c), we observed the (0.5 0.5 12.5) and (0.5 0.5 17.5) peaks characteristic of the $a^{-}$ and $b^{-}$ tilting \cite{glazer1975simple}. Based on their intensity ratio, we estimated the tilt angles ($\alpha$ and $\beta$) to be about 4$^{\circ}$. It is much larger than the single-layer SL but smaller than the bilayer SL \cite{supp,meyers2019magnetism}, indicating that the octahedral tilting is not the primary reason for the significant planar component in the bilayer slab of the HSL. Instead, it points to the fact that the HSL structure puts the two slabs of different confinement dimensions together in a unique environment, forming a new integrated system with properties distinct from each slab individually. Hypothetically, if the two slabs maintain the orthogonal moment directions, the inter-slab exchange coupling energy gain would vanish, making it difficult for either of them to stabilize long-range order due to the long-wavelength fluctuations at 2D \cite{mermin1966absence,coleman1973there}. In reality, they compromise with each other by adapting a primarily planar order in the bilayer and a lower $\it{T}$$_N$ for the monolayer.

In conclusion, we have engineered a HSL that combines single-layer and bilayer $J_{\rm eff} = 1/2$ square lattices which are known to have orthogonal anisotropy individually. Our systematic study shows that the HSL has an $a^{-}a^{-}c^{-}$ octahedral pattern and hosts a nearly planar canted AFM order through a single transition below 120 K. Compared to the single-layer and the bilayer SLs, the HSL stabilizes a new distinct state that cannot be described by the simple addition of the monolayer and bilayer properties since the proximity forces them to couple with each other electronically and magnetically. The results show that bringing monolayer and bilayer 2D systems with orthogonal properties close to each other in a hybrid superlattice structure is a powerful way to obtain unique states that cannot be achieved in a uniform structure, opening a way to search for new quantum states in layered materials.

When writing this paper, we notice a latest report on Sr$_5$Ir$_3$O$_{11}$ single crystal, which is effectively a (Sr$_2$IrO$_4$)$_1$/(Sr$_3$Ir$_2$O$_7$)$_1$ SL \cite{kim2022sr}. The single layer and bilayer therein are found to order separately at the same temperatures as Sr$_2$IrO$_4$ and Sr$_3$Ir$_2$O$_7$, respectively, with their AFM moments in the $ab$-plane and along the $c$-axis, respectively as well. Compared with the HSL, the inherent half-unit-cell slide of RP structure may significantly reduce the inter-slab coupling in Sr$_5$Ir$_3$O$_{11}$, preserving their individual properties.

The authors acknowledge experimental assistance from H. D. Zhou, H. Zhang and P. Siwakoti. J. L. acknowledges support from the National Science Foundation under Grant No. DMR-1848269. J. Y. acknowledges funding from the State of Tennessee and Tennessee Higher Education Commission (THEC) through their support of the Center for Materials Processing. L. Horak acknowledges support from the Czech Ministry of Education (project No. CZ.02.1.01/0.0/0.0/15\_003/0000485). Use of the Advanced Photon Source, an Office of Science User Facility operated for the U.S. DOE, OS by Argonne National Laboratory, was supported by the U.S. DOE under Contract No. DE-AC02-06CH11357.

\bibliography{SISTO}

\end{document}